\documentclass[journal]{IEEEtran}

\usepackage[left=0.75in, right=0.75in, top=1in, bottom=0.9in]{geometry}
\usepackage{mathrsfs,amsthm}
\usepackage{empheq,nccmath}

\usepackage{cite}

\makeatletter
\def\ps@headings{%
\def\@oddhead{\mbox{}\scriptsize\rightmark \hfil \thepage}%
\def\@evenhead{\scriptsize\thepage \hfil \leftmark\mbox{}}%
\def\@oddfoot{}%
\def\@evenfoot{}}
\makeatother 
\usepackage{graphicx,amsmath,amssymb,stfloats,subfigure,multicol}
\usepackage{epsfig,epsf,psfrag,amssymb,amsfonts,latexsym,graphicx,mathrsfs,subfigure}
\usepackage{bbm}
\usepackage{chngpage}
\usepackage[dvips]{color}
\usepackage{textcomp}
\usepackage{hyperref}
\usepackage{url}
\usepackage[hyphenbreaks]{breakurl}
\usepackage[normalem]{ulem}
\usepackage{algpseudocode}
\newsavebox{\ieeealgbox}

\usepackage{array}

\ifCLASSINFOpdf
\else
\fi
\newtheorem{theorem}{Theorem}

\newtheorem{corollary}{Corollary}
\newtheorem{lemma}{Lemma}
\newtheorem{definition}{Definition}

 \def\old#1{}    

\def\nn{\nonumber}
\def\beq{\begin{equation}}
\def\eeq{\end{equation}}
\def\bea{\begin{eqnarray}}
\def\eea{\end{eqnarray}}
\def\ba{\begin{array}}
\def\ea{\end{array}}

\def\bitem{\begin{itemize}}
\def\eitem{\end{itemize}}
\def\ben{\begin{enumerate}}
\def\een{\end{enumerate}}

\def\ie{{\it i.e.,\ \/}}

\definecolor{bgrd}{rgb}{1,1,1}
\definecolor{gray}{rgb}{0.5,0.5,0.5}
\definecolor{dkr}{rgb}{0.7,0.1,0.2}
\definecolor{dkb}{rgb}{0.1,0.1,0.8}

\makeatletter
\newdimen{\captionwidth}
\long\def\@makecaption#1#2{%
\captionwidth .9\hsize
\vskip 10pt%
\setbox\@tempboxa\hbox{#1: #2}%
  \ifdim \wd\@tempboxa >\captionwidth%
    \setbox\@tempboxa\hbox{#1:\hspace*{.5em}}%
    \hfil\parbox{\captionwidth}{\raggedright\hangindent \wd\@tempboxa%
    \hangafter=1\unhbox\@tempboxa#2}\hfill%
  \else\centerline{\box\@tempboxa}%
  \fi
}
\makeatother
\def\scalefig#1{\epsfxsize #1\textwidth}

\def\edoc{\end{document}}

\def\T{\mbox{\tiny T}}

\newcommand{\Hmsc}{\mathscr{H}}

\def\zetabf{\hbox{\boldmath$\zeta$\unboldmath}}
\def\etabf{\hbox{\boldmath$\eta$\unboldmath}}

\def\pibf{\hbox{\boldmath$\pi$\unboldmath}}

\def\psibf{\hbox{\boldmath$\psi$\unboldmath}}
\def\omegabf{\hbox{\boldmath$\omega$\unboldmath}}

\def\thetabf{{\mbox{\boldmath$\theta$\unboldmath}}}

\def\cbf{{\bf c}}
\def\dbf{{\bf d}}
\def\ebf{{\bf e}}
\def\fbf{{\bf f}}
\def\gbf{{\bf g}}

\def\pbf{{\bf p}}

\def\rbf{{\bf r}}

\def\rbf{{\bf r}}

\def\Abf{{\bf A}}

\def\Ebf{{\bf E}}

\def\Gbf{{\bf G}}

\def\Hc{{\cal H}}

\makeatletter
\let\old@ps@headings\ps@headings
\let\old@ps@IEEEtitlepagestyle\ps@IEEEtitlepagestyle
\def\psccfooter#1{%
    \def\ps@headings{%
        \old@ps@headings%
        \def\@oddfoot{\strut\hfill#1\hfill\strut}%
        \def\@evenfoot{\strut\hfill#1\hfill\strut}%
    }%
    \def\ps@IEEEtitlepagestyle{%
        \old@ps@IEEEtitlepagestyle%
        \def\@oddfoot{\strut\hfill#1\hfill\strut}%
        \def\@evenfoot{\strut\hfill#1\hfill\strut}%
    }%
    \ps@headings%
}
\makeatother

 \psccfooter{%
         \parbox{\textwidth}{\hrulefill \\ \small{22nd Power Systems Computation Conference} \hfill \begin{minipage}{0.2\textwidth}\centering \vspace*{4pt} \begin{psfrags}\scalefig{0.24}\epsfbox{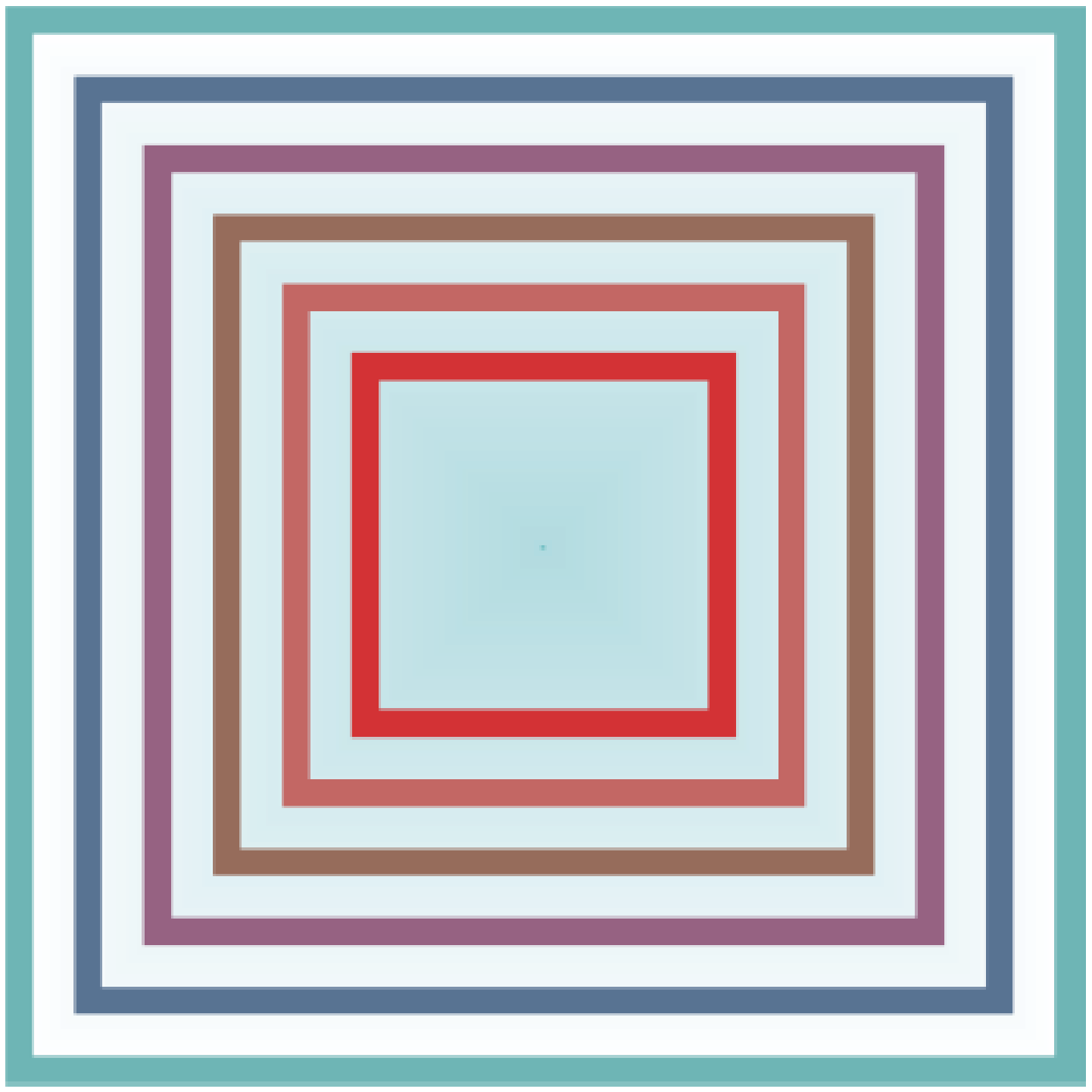}\end{psfrags}\\\small{PSCC 2022} \end{minipage} \hfill \small{Porto, Portugal --- June 27 -- July 1, 2022}}%
 }

\begin{document}


\title{Pricing Real-time Stochastic Storage Operations}
\author{
\IEEEauthorblockN{ \Large Cong Chen and Lang~Tong\\}%
\IEEEauthorblockA{School of Electrical and Computer Engineering\\
Cornell University, Ithaca, New York, USA\\
\{cc2662,lt35\}@cornell.edu}

\thanks{\scriptsize
\noindent Submitted to the 22nd Power Systems Computation Conference (PSCC 2022). This work is supported in part by the National Science Foundation under Award 1809830 and 1932501.Contact person: Lang Tong.}
}

\maketitle

\begin{abstract}
Pricing storage operation in the real-time market under demand and generation stochasticities is considered. A scenario-based stochastic rolling-window dispatch model is formulated for the real-time market, consisting of conventional generators, utility-scale storage, and distributed energy resource aggregators. We show that uniform pricing mechanisms require discriminative out-of-the-market uplifts, making settlements under locational marginal pricing (LMP) discriminative.   It is shown that the temporal locational marginal pricing (TLMP) that adds nonuniform shadow prices of ramping and state-of-charge to LMP
removes the need for out-of-the-market uplifts. Truthful bidding incentives are also established for price-taking participants under TLMP. Revenue adequacy and uplifts are evaluated in numerical simulations.
\end{abstract}

\begin{IEEEkeywords}
Stochastic storage operation, rolling-window look-ahead dispatch,  incentive compatibility, out-of-market uplifts,  locational marginal pricing.
\end{IEEEkeywords}

\section{Introduction}\label{sec:Intro}
We consider the problem of pricing storage operation within a rolling-window stochastic dispatch framework. This work is motivated by the increasing deployment of utility-scale storage and two recent landmark rulings by the United States Federal Energy Regulation Commission (FERC), FERC Order No. 841 and No. 2222, aimed at removing barriers to the participation of utility-scale storage and distributed energy resource aggregator (DERA) in capacity, energy, and ancillary service markets.

With the large-scale integration of renewable resources, power system operations face challenges arising from highly stochastic generation resources with strong temporal dependencies that often result in increasingly demanding ramping requirements. Meeting such ramping needs requires effective multi-interval dispatch and pricing mechanisms such as the flexible ramping products introduced in the real-time market by multiple system operators \cite{CAISO_FRP:15,Wang&etal:16FERC}.  With the broadening participation of utility-scale electric storage resource (ESR) and DERA in wholesale electricity markets, there are compelling needs for effective multi-interval market operations.

A widely adopted approach to multi-interval operations in the real-time market is the rolling-window dispatch, where the operator sequentially optimizes the immediate dispatch decision based on forecasted demands and supplies over a finite look-ahead window. Such an approach provides a computationally tractable data-driven solution, exploiting that forecasts are more accurate for intervals close to the dispatch time.

Pricing multi-interval dispatch under uncertainty, however, faces a different set of challenges arising from inevitable forecast errors that directly affect dispatch and pricing decisions.  It has been shown recently \cite{Guo&Chen&Tong:21TPS,Chen&Guo&Tong:20TPS,ChenTongGuo:21pricing} that all uniform pricing schemes for rolling-window dispatch result in lost opportunity costs (LOCs) that require discriminative uplift payments in a non-transparent out-of-the-market settlement process. With the increasing participation of utility-scale ESRs and DERAs, LOC arises not only from generation ramping constraints but also from binding constraints on states of charge (SOC).   Because  LOC payments are computed based on  bid-in costs, ramping parameters, and SOC parameters, there are incentives for generators, ESRs, and DERAs to manipulate bidding parameters, e.g., withholding ramping and SOC limits, to gain additional profit from LOC uplifts. See examples in Sec.~ \ref{sec:UntruBidExample}.
%
%

\subsection{Summary of results and contributions}
We consider the problem of pricing ESR operations in the real-time energy market under a {\em stochastic rolling window dispatch optimization} with a generalized ESR model that includes conventional generators and DERA.

The main contribution of this work is threefold.  First, we establish that all uniform pricing schemes such as LMP result in LOCs, making discriminative out-of-the-market uplifts necessary to provide dispatch-following incentives.  The significance of this result is that discriminative settlement is unavoidable for the stochastic rolling-window dispatch.

Second, we extend LMP to temporal locational marginal pricing (TLMP) by adding nonuniform shadow prices associated with ramping and state-of-charge (SOC) constraints.  We show that, in contrast to LMP, TLMP for ESRs results in zero LOC regardless of the accuracy of demand and stochastic generation forecasts. This is surprising, perhaps, because LOC is computed ex-post after uncertainties are realized, whereas TLMP is calculated in real-time involving inaccurate forecasts and tentative dispatch in future intervals.

Third, we examine truthful-bidding incentives under rolling-window stochastic dispatch with LMP and TLMP.  We show that, if an ESR bids under a price-taking assumption, it is (locally) optimal to bid truthfully with its capacity, SOC limits, and ramp limits under TLMP.  Under LMP, in contrast, it is not optimal to bid truthfully because of the possibility of manipulating bidding parameters to gain profits from out-of-the-market uplift payments.

Finally, we provide a set of numerical results that compare LMP and TLMP on a broader set of performance metrics.  In particular, we examine the impacts of ESR and the use of stochastic rolling-window dispatch on uplift payments (under LMP) to generators and ESRs, and the merchandising surplus of ISO. We demonstrate cases that stochastic rolling-window dispatch has lower uplift payments (under LMP) than that under the deterministic dispatch model. Our results also show a positive merchandising surplus for the system operator under TLMP, for which the operator may redistribute the surplus among consumers and energy resources.

\subsection{Related work}
Stochastic rolling-window dispatch under forecasting uncertainties has been widely studied \cite{Eleni20EPSR,Emmanouil18EPSR,Harsha16TPS,JiadongWang13,PatrinosBemporad11CDC,AndySun15TPS,LeXie14TSG,Nazir20EPSR}. Various stochastic models have been implemented in rolling-window dispatch when considering uncertainties from renewables and demands, including scenario-based stochastic optimization \cite{Eleni20EPSR,Emmanouil18EPSR,Harsha16TPS,JiadongWang13,PatrinosBemporad11CDC}, robust optimization\cite{AndySun15TPS,LeXie14TSG}, and chance-constrained stochastic optimization \cite{Nazir20EPSR}. These works highlight the advantages of incorporating uncertainty models in rolling-window dispatch over a conventional deterministic approaches. In particular, there has been considerable evidence that stochastic rolling-window dispatch can reduce operation costs \cite{Emmanouil18EPSR,Harsha16TPS,AndySun15TPS,JiadongWang13,PatrinosBemporad11CDC}  and produce reliable scheduling plans \cite{Eleni20EPSR,Nazir20EPSR}.

To our best knowledge, there is no published work on pricing real-time ESR operations under stochastic rolling-window dispatch models with the exception of \cite{philpott2016equilibrium, WangHobbs14EPSR} .  In \cite{philpott2016equilibrium}, competitive equilibrium conditions are established for multi-stage stochastic operation with hydro-electric reservoirs under LMP. And in \cite{WangHobbs14EPSR}, the scenario-based stochastic rolling window dispatch with LMP was derived as an ideal formulation to be approximated by a deterministic model with a flexible ramp product.

The problem of pricing deterministic rolling-window dispatch, however, has attracted considerable interest recently  \cite{hogan2020Multi-intervalPricing,Hua&etal:19TPS,Zhao&Zheng&Litvinov:19TPS,Guo&Chen&Tong:21TPS,Chen&Guo&Tong:20TPS,ChenTongGuo:21pricing}; much of these recent works focus on the issue of dispatch-following incentives and the need for out-of-the-market uplifts. The idea of generalizing LMP to nonuniform temporal locational marginal pricing (TLMP) is first proposed  in  \cite{Guo&Chen&Tong:21TPS,Chen&Guo&Tong:20TPS,ChenTongGuo:21pricing}, where it's shown that, under reasonable operating scenarios, all uniform pricing schemes result in LOC and require out-of-the-market uplifts. This result seeds the parallel result under the stochastic rolling-window model presented in Theorem~\ref{thm:StorageUniformPricingLOC}  \& \ref{thm:R-TLMPLOC} in Sec.~\ref{sec:IC}.

It has been shown in \cite{Guo&Chen&Tong:21TPS,Chen&Guo&Tong:20TPS,ChenTongGuo:21pricing}  that TLMP removes LOC independent of forecasting error used in the rolling-window dispatch, and it is optimal for a price-taking generator to bid with its true marginal cost.  The authors also validate numerically that, under TLMP, it is optimal for a generator to reveal truthfully its ramp limits.  The incentive to truthfully reveal bidding parameters under TLMP in the rolling-window dispatch is established here formally by Theorem.~\ref{thm:TLMPBidLOCESR} in Sec.~\ref{sec:IC}. This result is in parallel to the earlier work of Gross and Finlay \cite{gross2000generation} that shows the truthful revelation behavior in a perfectly competitive market under LMP in the deterministic one-shot dispatch. 




\section{Stochastic Rolling-window Dispatch}\label{sec:Model}
We formulate in this section a stochastic rolling-window dispatch model consisting of a generalized ESR model, a stochastic demand and its probabilistic forecasts, and a scenario-based stochastic optimization.  

\subsection{A general ESR Model for dispatchable resources}
We define a generalized ESR model to include ESR,  DERA as virtual storage unit,  and conventional generator/elastic demand as storage with specific parameters.

Let $\Hmsc=\{1, 2, \cdots, T\}$ be the set of all dispatch intervals and $\Hmsc_t=\{t, \cdots, t+W-1\}$  the dispatch intervals of the $W$-interval rolling-window starting at interval $t$. For ESRs, we use superscript ``D'' for discharging and ``C''  for charging.  In interval $t$, decision variables associate with ESR unit $i$ are denoted by
$(g^{\mbox{\tiny D}}_{it},g^{\mbox{\tiny C}}_{it},E_{it})$, where  $E_{it}$ is the SOC and $g^{\mbox{\tiny D}}_{it},g^{\mbox{\tiny C}}_{it}$ the discharging and charging decision variables, respectively. Parameters of ESR $i$ are defined by storage boundary limits $(\underline{E}_i, \bar{E}_i)$,
charging/discharging power limits $(\underline{g}_i^{\mbox{\tiny D}},\bar{g}_i^{\mbox{\tiny D}},\underline{g}_i^{\mbox{\tiny C}},\bar{g}_i^{\mbox{\tiny C}})$, and  the up and down ramp limits $(\underline{r}_i^{\mbox{\tiny D}},\bar{r}_i^{\mbox{\tiny D}},\underline{r}_i^{\mbox{\tiny C}},\bar{r}_i^{\mbox{\tiny C}})$.

The ESR definition above includes conventional generator and DERA as special cases.  For example, by setting the storage boundary limits $\bar{E} = \infty$, $\underline{E}=-\infty$, and discharging limits to zero, we obtained the standard generation model. A DERA can be modeled as a virtual storage without SOC. 

\subsection{Stochastic demand,  forecasts, and decision variables}
Let  $(d_t)$ be a realization of stochastic inelastic demand process.  We assume that probabilistic demand forecasts are made available at time $t$ over the current $W$-interval look ahead window $\Hmsc_t$ in the form of $K$  possible scenarios $\{\hat{\dbf}_{tk}, k=1,\cdots,K\}$, where $\hat{\dbf}_{tk}=(\hat{d}_{tk},\cdots, \hat{d}_{(t+W-1)k})$ is the $k$th scenario of forecasted demand over $\Hmsc_t$. We assume that demand forecasts are perfect in interval $t$, \ie $\hat{d}_{tk}=d_t$. Note that demand forecasts are updated every time when the rolling window move forward at time $t$. 

 Let $N$ be the number of ESRs in a single-bus model. The decision variables in dispatch window $\Hmsc_t$ include charging/discharging variables in the binding interval $t$  and $K$ charging/discharging variables in the advisory intervals under each demand scenario.
   For ESR unit $i$, let  $(g_{it}^{\mbox{\tiny C}}, g_{it}^{\mbox{\tiny D}})$ be the charging/discharging decision variables in interval $t$ and $\gbf[t]=(\gbf^{\mbox{\tiny C}}[t],\gbf^{\mbox{\tiny D}}[t])$ the charging/discharging vectors involving all ESR units. For the advisory interval $t'$, let $\gbf_k[t]=(\gbf_k^{\mbox{\tiny C}}[t],\gbf_k^{\mbox{\tiny D}}[t])$ be the decisions under scenario $k$.   
   
   Finally, we collect all the decision variables in
  $\Gbf_t$ that includes decision vectors in the binding interval and the advisory intervals for all scenarios.  Similarly defined is the matrix $\Ebf_t$ that includes the state-of-the-charge variables associated with $\Gbf_t$.


\subsection{Stochastic rolling window dispatch}
\begin{figure}[h]
 \center
 \begin{psfrags}
 \scalefig{0.5}\epsfbox{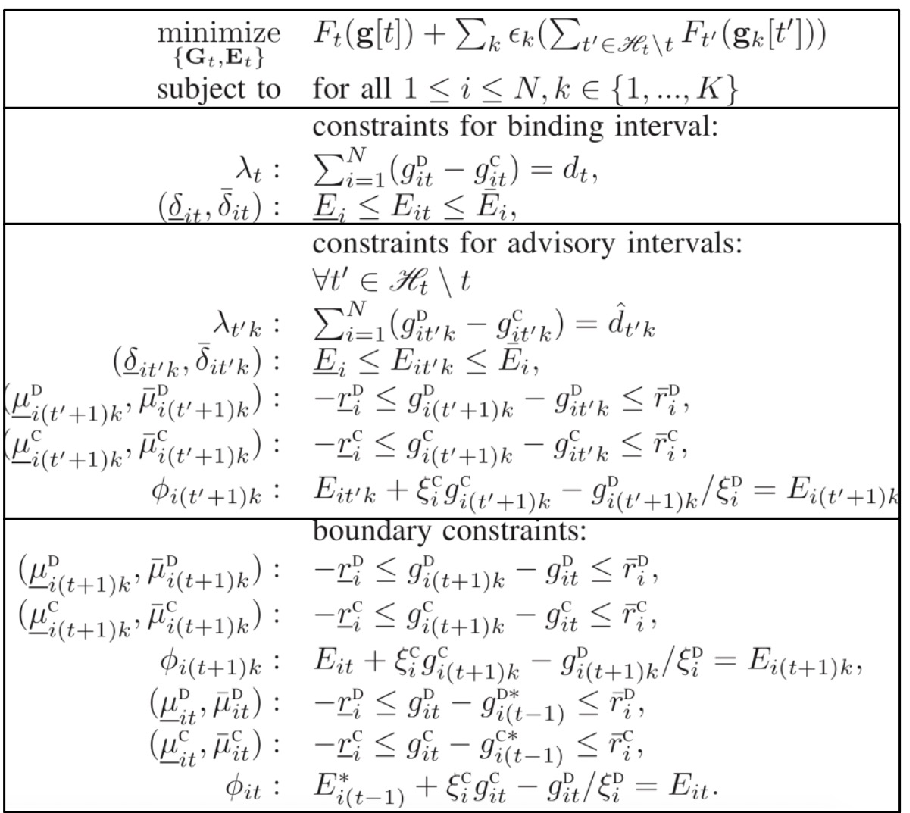}
 \end{psfrags}
\vspace{-1em}
 \caption{\small Stochastic rolling-window dispatch problem.}
 \label{fig:StochasticOptimization}
 \end{figure}

The stochastic optimization in the dispatch window $\Hmsc_t$ is given in Fig.~\ref{fig:StochasticOptimization} with the defined generalized ESR model representing various dispatchable resources. We summarize here key aspects of this optimization.

The dispatch cost for the binding interval $t$ is given by
\beq
F_t(\gbf[t]):=\sum_{i=1}^N (f^{\mbox{\tiny D}}_{it} (g^{\mbox{\tiny D}}_{it})- f^{\mbox{\tiny C}}_{it} (g^{\mbox{\tiny C}}_{it})) 
\eeq
where $\{f_{it}^{\mbox{\tiny C}}\}$ and $\{f_{it}^{\mbox{\tiny D}}\}$ are bid-in charging and discharging costs, respectively.  The total {\em expected cost}  in the dispatch window $\Hmsc_t$ is then given by
\beq
\bar{F}_t(\Gbf_t)=F_t(\gbf[t]) + \sum_k \epsilon_k (\sum_{t'\in \Hmsc_t \setminus t}   F_{t'}(\gbf_{k}[t']) ),
\eeq
where $\epsilon_k$ represents probability of the $k$-th scenario. And in Fig.~\ref{fig:StochasticOptimization}, for each forecasted demand scenario $\hat{\dbf}_{tk}$, power balance equations and a set of dispatch constraints associated with the forecasted demands are defined.  

Dual variables most relevant in defining prices in the binding interval $t$ are $\lambda_t$ associated with the power balance constraint in $t$, the ramping shadow prices for each scenario $(\bar{\mu}_{itk}^{\mbox{\tiny C}}, \underline{\mu}_{itk}^{\mbox{\tiny C}}, \bar{\mu}_{itk}^{\mbox{\tiny D}}, \underline{\mu}_{itk}^{\mbox{\tiny  D}})$, the SOC shadow prices $(\phi_{it})$, and the ramping shadow prices for initial boundary conditions $(\bar{\mu}_{it}^{\mbox{\tiny C}}, \underline{\mu}_{it}^{\mbox{\tiny C}}, \bar{\mu}_{it}^{\mbox{\tiny D}}, \underline{\mu}_{it}^{\mbox{\tiny  D}})$\footnote{It's assumed throughout the paper that all dual variables for inequality constraints are defined in a way to be always non-negative.}.

 We use superscript ``*'' to indicate solutions to Fig.~\ref{fig:StochasticOptimization}.  In particular, let $(g_{it}^{{\mbox{\tiny D}}*},g_{it}^{{\mbox{\tiny C}}*}, E^*_{it})$ be the solution of the stochastic dispatch in the binding interval. In Fig.~\ref{fig:StochasticOptimization}, $\xi_i^{\mbox{\tiny D}} ,\xi_i^{\mbox{\tiny C}} \in (0,1]$ are charging and discharging efficiency coefficients of ESR $i$. By assuming $\frac{\partial}{\partial g^{\mbox{\tiny D}}_{it}} f^{\mbox{\tiny D}}_{it} (g^{\mbox{\tiny D}}_{it}) > \frac{\partial}{\partial g^{\mbox{\tiny C}}_{it}} f^{{\mbox{\tiny C}}}_{it} (g^{\mbox{\tiny C}}_{it})/\xi_i,\forall i, \forall t \in \Hmsc_t$ with $\xi_i=\xi_i^{\mbox{\tiny D}}\xi_i^{\mbox{\tiny C}}$, constraints like $ g^{\mbox{\tiny D}}_{it}g^{\mbox{\tiny C}}_{it}=0$ can be exactly relaxed in most cases \cite{ZLi2018sufficientStorageRelax}. In boundary constraints, initial conditions for the realized dispatch $(\gbf^{\mbox{\tiny D}*}[t-1],\gbf^{\mbox{\tiny C}*}[t-1])$, and SOC in interval $t-1$, $\Ebf^{*}[t-1]$, follow the optimal binding solution at the $(t-1)$-th rolling window.

We present two marginal cost pricing schemes under the stochastic rolling-window dispatch model; one is the standard uniform locational marginal price (LMP), the other the non-uniform extension of LMP---TLMP.  In both cases, the prices in interval $t$ are determined in the dispatch window $\Hmsc_t$ given the realizations of demand $d_t$ and forecast scenarios $(\hat{d}_{tk})$.

\subsection{Locational Marginal Pricing}
LMP is defined by the marginal cost of meeting the demand $d_t$  for given forecasted future demand scenarios. By the envelope theorem, 
\beq \label{eq:RWLMP}
\pi^{\mbox{\rm\tiny LMP}}_{t} := \frac{\partial}{\partial d_t} \bar{F}_t(\Gbf^*_t) = \lambda_t^*,
\eeq
where $\lambda_{t}^{*}$ is the solution of the stochastic optimization in Fig.~\ref{fig:StochasticOptimization}.  All resources have the same price $\pi^{\mbox{\rm\tiny LMP}}_{t}$ in interval $t$.

\subsection{Temporal Locational Marginal Pricing}
TLMP is a non-uniform marginal cost pricing that measures the marginal contribution of the resource to meeting the demand $d_t$ at the optimal dispatch. Let
$\bar{F}^{(-i)}_t(\Gbf_t^*)$ be the total cost in rolling-window $t$, excluding contribution of ESR $i$ in interval $t$  by treating $(g_{it}^{\mbox{\tiny C *}},g_{it}^{\mbox{\tiny D *}})$  as parameters set at the optimal dispatch point, i.e.   
\beq
\begin{array}{lrl}
\bar{F}^{(-i)}_t(\Gbf_t^*)&=&\bar{F}_t(\Gbf_t^*)-(f^{\mbox{\tiny D}}_{it} (g^{\mbox{\tiny D *}}_{it})- f^{\mbox{\tiny C}}_{it} (g^{\mbox{\tiny C *}}_{it})).
\end{array}
\eeq
The marginal contribution from ESR $i$ in interval $t$ is given by
\bea \label{eq:RTLMPESR}
\pi_{it}^{\mbox{\tiny TLMP}}&:=&
 \left\{\begin{array}{ll}
 \frac{\partial}{\partial g^{\mbox{\tiny C}}_{it}}  \bar{F}^{(-i)}_t(\Gbf^*_t) & \mbox{ESR charging}\\
 - \frac{\partial}{\partial g^{\mbox{\tiny D}}_{it}}  \bar{F}^{(-i)}_t(\Gbf^*_t) & \mbox{ESR discharging}\\
 \end{array}\right.\\
&=&
  \left\{\begin{array}{ll}
 \lambda^*_t- \xi_i^{\mbox{\tiny C}}\phi^*_{it}-\Delta_{it}^{\mbox{\tiny C}*} :=\pi_{it}^{\mbox{\tiny TLMP-C}}, & \mbox{charging,}\\
 \lambda^*_t- 1/\xi_i^{\mbox{\tiny D}}\phi^*_{it}+\Delta_{it}^{\mbox{\tiny D}*}:=\pi_{it}^{\mbox{\tiny TLMP-D}}, & \mbox{discharging,}\\
    \end{array}
    \right.
\eea
where $\Delta_{it}^{\mbox{\tiny C}*}$ (and, similarly,  $\Delta_{it}^{\mbox{\tiny D}*}$) is  defined by
\bea
\Delta_{it}^{\mbox{\tiny C}*}&:=&-(\bar{\mu}_{it}^{\mbox{\tiny C}*}-\underline{\mu}_{it}^{\mbox{\tiny C}*}) +  \sum_{ k =1}^K \bigg(\bar{\mu}_{i(t+1)k}^{\mbox{\tiny C}*}-\underline{\mu}_{i(t+1)k}^{\mbox{\tiny C}*}\bigg).
\eea
TLMP prices the inelastic demand with the same way as LMP. Note that TLMP $\pi_{it}^{\mbox{\tiny TLMP}}$ for ESR $i$ can be decomposed into the energy price $\pi^{\mbox{\rm\tiny LMP}}_{t}=\lambda^*_t$, the SOC price $\phi^*_{it}$, and ramping prices $(\Delta_{it}^{\mbox{\tiny C}*}, \Delta_{it}^{\mbox{\tiny D}*})$. Dual values used for LMP and TLMP come from the optimal dual solution of stochastic rolling-window dispatch in Fig.~\ref{fig:StochasticOptimization}. 

The following lemma provides another formulation for the SOC price using dual variables for constraints of SOC limits rather than SOC transition equations.
\begin{lemma}[SOC price]\label{lemma:SOC}
With the optimal dual solution of stochastic rolling-window dispatch in Fig.~\ref{fig:StochasticOptimization}, the SOC price has 
\[\phi^*_{it}=\Delta \delta_{it}^*+\sum_{t'=t+1}^{t+W-1}\sum_{k=1}^K\Delta \delta_{it'k}^*,\]
where $\Delta \delta_{it}^*:=\underline{\delta}_{it}^*-\bar{\delta}_{it}^*, \Delta \delta_{it'k}^*:=\underline{\delta}_{it'k}^*-\bar{\delta}_{it'k}^*$.
\end{lemma}
{\em Proof:}  See the appendix for the proof.  

Although it's not obvious when SOC price $\phi^*_{it}$ , the dual variable of the SOC transition equation, equals zero, Lemma~\ref{lemma:SOC} indicates that  $\phi^*_{it}=0$ when there is no binding SOC limit constraint. We further explore the case when TLMP reduces to LMP in the following corollary.
\begin{corollary}\label{corr:TLMP=LMP}
TLMP reduces to LMP for storage $i$ at time $t$, \ie $\pi_{it}^{\mbox{\tiny TLMP}}=\pi^{\mbox{\rm\tiny LMP}}_{t} $, if the following conditions are satisfied in the $W$-interval look ahead window $\Hmsc_t$ :
\ben
\item there's no binding ramping constraint for storage $i$ from  $t-1$ to $t$, and from $t$ to $t+1$.
\item there's no binding SOC limit constraint  for storage $i$ from $t$ to $t+W-1$ .
\een
\end{corollary}
{\em Proof:}  See the appendix for the proof. In general, TLMP, as an extension of LMP, reduces to LMP when there are no binding ramping and SOC limit constraints.

\section{Incentive compatibility}\label{sec:IC}

We analyze the incentive compatibilities of LMP and TLMP under the stochastic rolling-window dispatch model with two types of incentives.

The first type is the dispatch following incentive that addresses the issue of whether a profit-maximizing  ESR would produce the same dispatch by itself in ex-post, given the realized price sequence.  A pricing scheme is compatible with the dispatch-following incentive if  every ESR has the same profit as that from the ESR's individual profit optimization using the realized price sequence over $\Hmsc$.  Therefore, a necessary and sufficient condition for dispatch-following incentive compatibility is that no ESR requires LOC uplifts.

Note that, under the stochastic model of inelastic demand,  the dispatch signal $(\gbf^{\mbox{\tiny D}*}, \gbf^{\mbox{\tiny C}*})$  are random, which makes LOC of an ESR a random variable.   To this end, dispatch-following incentive compatibility of a pricing scheme for a stochastic rolling-window dispatch is defined as follows.

\begin{definition} A pricing scheme $\pibf$ is compatible with dispatch-following incentives if, with probability one,  the LOCs of all ESRs are zero.
\end{definition}

A second type is the truthful-bidding incentive. A pricing scheme is compatible with truthful-bidding incentive if it is optimal for every ESR to reveal its true cost function and bidding parameters such as the generation, ramp, and SOC limits.  To this end, it is necessary to show that, given the realized price sequence, profit maximization using the true parameters is locally optimal.  In other words, no ESR will deviate from its the truthful bidding action in revealing the actual costs and bidding parameters.

\subsection{LOC as a measure of dispatch-following incentives}
The dispatch-following incentives of a pricing scheme can be measured by the lost opportunity costs (LOC) if the ESR follows the operator's dispatch signal\footnote{The ESR index $i$ is dropped in (\ref{eq:uplift})(\ref{eq:Q}) for brevity} .   

Let $(\gbf^{\mbox{\tiny C}*}, \gbf^{\mbox{\tiny D}*})$ be the column vectors of charging and discharging dispatch.  Let $\pibf=(\pi_1,\cdots,\pi_T)$ be the column vector of a realized  uniform price over the entire scheduling horizon.  The LOC under $\pibf$  is defined by the difference between
the maximum profit achievable under the individual optimization given the ex-post price $\pibf$ and the profit realized by the operator's dispatch:  
\beq \label{eq:uplift}
\begin{array}{c}
{\rm LOC}(\pibf,\gbf^{\mbox{\tiny C}*},\gbf^{\mbox{\tiny D}*})
=Q(\pibf) - \bigg(\pibf^{\T}\gbf^{\mbox{\tiny D}*}-\pibf^{\T}\gbf^{\mbox{\tiny C}*}\\
~~~~-\sum_{t=1}^T f^{\mbox{\tiny D}}_{t}(g^{\mbox{\tiny D}*}_{t})+\sum_{t=1}^T f^{\mbox{\tiny C}}_{t}(g^{\mbox{\tiny C}*}_{t})\bigg),
\end{array}
\eeq
where the terms in the parentheses compute the profit realized from the settlement within the real-time market, and 
$Q(\pibf)$ is the maximum profit the ESR would have received under $\pibf$ through the individual profit maximization defined by
\beq \label{eq:Q}
\begin{array}{lcl}
&Q(\pibf)= \underset{\{\pbf^{\mbox{\tiny D}},\pbf^{\mbox{\tiny C}},\ebf\}}{\rm maximize} &
 \pibf^{\T}\pbf^{\mbox{\tiny D}}-\pibf^{\T}\pbf^{\mbox{\tiny C}}\\
&&+\sum_{t=1}^T \bigg(f^{\mbox{\tiny C}}_{t}(p^{\mbox{\tiny C}}_{t}) - f^{\mbox{\tiny D}}_{t}(p^{\mbox{\tiny D}}_{t})\bigg)\\
& {\rm subject~to}&\\
&\psibf: & \xi^{\mbox{\tiny C}}\pbf^{\mbox{\tiny C}}-\pbf^{\mbox{\tiny D}}/\xi^{\mbox{\tiny D}}=\Abf\ebf, \\
& (\underline{\etabf}_{i}^{\mbox{\tiny D}},\bar{\etabf}_{i}^{\mbox{\tiny D}}):  & -\underline{\rbf}_i^{\mbox{\tiny D}}\le \Abf \pbf_i^{\mbox{\tiny D}} \le \bar{\rbf}_{i}^{\mbox{\tiny D}}, \\
& (\underline{\etabf}_{i}^{\mbox{\tiny C}},\bar{\etabf}_{i}^{\mbox{\tiny C}}):  & -\underline{\rbf}_i^{\mbox{\tiny C}}\le \Abf \pbf_i^{\mbox{\tiny C}} \le \bar{\rbf}_{i}^{\mbox{\tiny C}}, \\
&(\underline{ \omegabf },\bar{ \omegabf }): & \underline{\Ebf}\le \ebf \le \bar{\Ebf}, \\
 &(\underline{\zetabf}^{\mbox{\tiny D}},\bar{\zetabf}^{\mbox{\tiny D}}):&{\bf 0}\leq \pbf^{\mbox{\tiny D}} \leq\bar{\gbf}^{\mbox{\tiny D}},\\
 &(\underline{\zetabf}^{\mbox{\tiny C}},\bar{\zetabf}^{\mbox{\tiny C}}):&{\bf 0}\leq \pbf^{\mbox{\tiny C}} \leq \bar{\gbf}^{\mbox{\tiny C}},
\end{array} \hfill
\eeq
where  $\pbf^{\mbox{\tiny D}},\pbf^{\mbox{\tiny C}},\ebf$ are, respectively, decision variables for  discharging power, charging power and SOC, and $\Abf$, a  $T \times T$ lower bidiagonal matrix with 1 as diagonals and $-1$ left next to diagonals, defines the SOC evolution.

\subsection{Dispatch-following incentive compatibility}
We establish dispatch-following incentive properties for LMP and TLMP.  To this end, we generalize a result in \cite{Guo&Chen&Tong:21TPS,Chen&Guo&Tong:20TPS,ChenTongGuo:21pricing}  for the  stochastic rolling-window dispatch model.

The first result shows that, under conditions that hold with a positive probability, no uniform pricing scheme is compatible with dispatch-following incentives.

\begin{theorem}[Dispatch-following incentive compatibility of uniform pricing] \label{thm:StorageUniformPricingLOC}
Assume the random inelastic demand and probabilistic load forecasts have continuous distributions. A uniform pricing scheme under the  stochastic rolling-window dispatch is not compatible with dispatch-following incentives if there exist ESR $i$ and $j$ satisfying the following conditions:
\ben
\item ESR $i$ and  $j$ have different charging and discharging bid curves;
\item there exists realizations of  the dispatch $(g_{it}^{\mbox{\tiny C}*},g_{it}^{\mbox{\tiny D}*})$ and  $(g_{jt}^{\mbox{\tiny C}*},g_{jt}^{\mbox{\tiny D}*})$  such that both ESRs are ``marginal'' in some interval  $t^* \in \Hmsc$, \ie
\ben
    \item both ESRs do not reach charging or discharging limits in $t^*$.
    \item both ESRs have no binding ramping constraints from intervals $t^*-1$ to $t^*$ and from $t^*$ to $t^*+1$.
    \item both ESRs do not reach SOC limits from $t^*$ to the end of $\Hmsc$.
\een
 \een
\end{theorem}
{\em Proof:} See the appendix for the proof.

An analysis of the probability for which the conditions above hold appears to be intractable.  In our simulations using typical load profiles, these conditions hold with fairly high probabilities.  See  \cite{Chen&Guo&Tong:20TPS, ChenTongGuo:21pricing}.

The second result shows that, as a generalization of LMP, TLMP is compatible with dispatch-following incentives.

\begin{theorem}[Dispatch-following incentive compatibility of TLMP] \label{thm:R-TLMPLOC}
 \label{thm:TLMPBidLOCESR} 
 For ESR $i$, let $\gbf_i^{\mbox{\tiny D}*},\gbf_i^{ \mbox{\tiny C}*}$ be the rolling-window dispatch over $\Hmsc$  computed from Fig.~\ref{fig:StochasticOptimization}, and  $\pibf_i^{\mbox{\rm \tiny TLMP}}$ its TLMP sequence.  Then, for all $i$ and under all realizations of stochastic demands and probabilistic load forecasts
\beq \label{eq:StorageLOC=0}
\mbox{\rm LOC}(\pibf_i^{\mbox{\rm \tiny TLMP}},\gbf_i^{\mbox{\tiny D}*},\gbf_i^{\mbox{\tiny C}*})=0.
\eeq
\end{theorem}

{\em Proof:}   See the appendix for the proof.

\subsection{Truthful-bidding incentive compatibility}
We now consider truthful-bidding incentives, by which we mean that it is optimal for a profit-maximizing ESR to reveal truthfully its marginal cost and bidding parameters.  

For ESR $i$, we parameterize the charging/discharging bids  $(\fbf_i^{\mbox{\tiny C}}(\gbf_i^{\mbox{\tiny C}}|\thetabf_i),\fbf_i^{\mbox{\tiny D}}(\gbf_i^{\mbox{\tiny D}}|\thetabf_i))$  by $\thetabf_i$, where 
 $\fbf_i^{\mbox{\tiny C}}(\gbf_i^{\mbox{\tiny C}}|\thetabf_i) = (f_{it}^{\mbox{\tiny C}}(g_{it}^{\mbox{\tiny C}}|\thetabf_i))$, $\fbf_i^{\mbox{\tiny D}}(\gbf_i^{\mbox{\tiny D}}|\thetabf_i) = (f_{it}^{\mbox{\tiny D}}(g_{it}^{\mbox{\tiny D}}|\thetabf_i))$, and they
are the vectors of charging/discharging bid curves over the scheduling horizon $\Hc$.  The bidding parameter $\thetabf_i$ includes the bid-in marginal cost and operational characteristics such as ramp limits, generation capacity,  and state-of-charge limits.

We make the assumption that ESRs are price-takers in the sense that the bids and offers from ESRs are constructed under the assumption that their bids cannot influence the market clearing price.  Specifically, under TLMP,  in constructing its bid, a price-taking  ESR $i$   maximizes its profit of the form 
\bea \label{eq:StoragePi_i}
 \Pi_i^{\mbox{\tiny TLMP}}(\thetabf_i) &=& \sum_{t=1}^T \bigg(\pi_{it}^{\mbox{\tiny TLMP-D}}g_{it}^{\mbox{\tiny D}*}(\thetabf_i)-\pi_{it}^{\mbox{\tiny TLMP-C}}g_{it}^{\mbox{\tiny C}*}(\thetabf_i)\nn\\
 & & -q_{it}^{\mbox{\tiny D}}(g^{\mbox{\tiny D}*}_{it}(\thetabf_i))+q_{it}^{\mbox{\tiny C}}(g^{\mbox{\tiny C}*}_{it}(\thetabf_i))\bigg)
\eea
where, $(g_{it}^{\mbox{\tiny C}*}(\thetabf_i),g_{it}^{\mbox{\tiny D}*}(\thetabf_i))$  are the charging/discharging dispatch signals from the market clearing process, and 
$(q_{it}^{\mbox{\tiny C}}(\cdot), q_{it}^{\mbox{\tiny D}}(\cdot))$ are the true benefit-cost curves of charging and discharging, respectively.  Note with price-taking assumption the clearing prices are not functions of bidding parameter $\thetabf_i$.

Under LMP, on the other hand, the profit of an ESR includes not only the in-market credits and charges but also the out-of-the-market uplifts:
\bea \label{eq:StoragePi_i_LMP}
 &\Pi^{\mbox{\tiny LMP}}_i(\thetabf_i) =\sum_{t=1}^T \bigg(\pi_{t}^{\mbox{\tiny LMP}}(g_{it}^{\mbox{\tiny D}*}(\thetabf_i)-g_{it}^{\mbox{\tiny C}*}(\thetabf_i))
 -q_{it}^{\mbox{\tiny D}}(g^{\mbox{\tiny D}*}(\thetabf_i)) +\nn\\
 &~~q_{it}^{\mbox{\tiny C}}(g^{\mbox{\tiny C}*}(\thetabf_i))\bigg) +{\rm LOC}(\pibf^{\mbox{\tiny LMP}}, \gbf_i^{\mbox{\tiny D}*}(\thetabf_i),\gbf_i^{\mbox{\tiny C}*}(\thetabf_i))
\eea
where, from (\ref{eq:uplift}), the LOC term depends on $\thetabf_i$, making it possible for ESR $i$ to manipulate $\thetabf_i$ to optimize LOC payment and the overall profit.  It is shown in Sec.~\ref{sec:UntruBidExample} that,  under LMP,  a generator can make more profits by withholding maximum ramp limit.  

The following theorem shows the truthful-biding incentive compatibility of TLMP.

\begin{theorem}[Truthful-bidding incentive compatibility of TLMP] \label{thm:TLMPBidLOCESR}  Under TLMP and stochastic rolling-window dispatch, for all realizations of stochastic demands and probabilistic load forecasts, it is optimal for price-taking ESRs to bid  with its true benefit/cost curves and true operational parameters.
\end{theorem}
{\em Proof:}  See the appendix for the proof. Note that the price-taking assumption does not imply that bids constructed under such an assumption cannot affect the market clearing prices. Specifically, an ESR that constructs its bid based on the price-taking assumption may become a marginal generator under specific demand and generation conditions.  In such a case, an ESR can indeed influence the market clearing price.  In practice, however, an ESR without market power cannot  foresee exactly when it may become a marginal
 generator.  Thus it is optimal for all practical purposes to bid truthfully.


\section{Case Studies}\label{sec:CaseStudies}
We present four Monte Carlo case studies on a single bus network with three generators.  Case 1 involves no ESR and uses deterministic rolling-window dispatch.   Case 2 involves no ESR and uses stochastic rolling-window dispatch.  Case 3 involves one ESR and uses deterministic rolling-window dispatch.  Case 4 involves one ESR and uses stochastic rolling-window dispatch. 

\subsection{Simulation settings}
\begin{figure}[h]
\center
\begin{psfrags}
\scalefig{0.25}\epsfbox{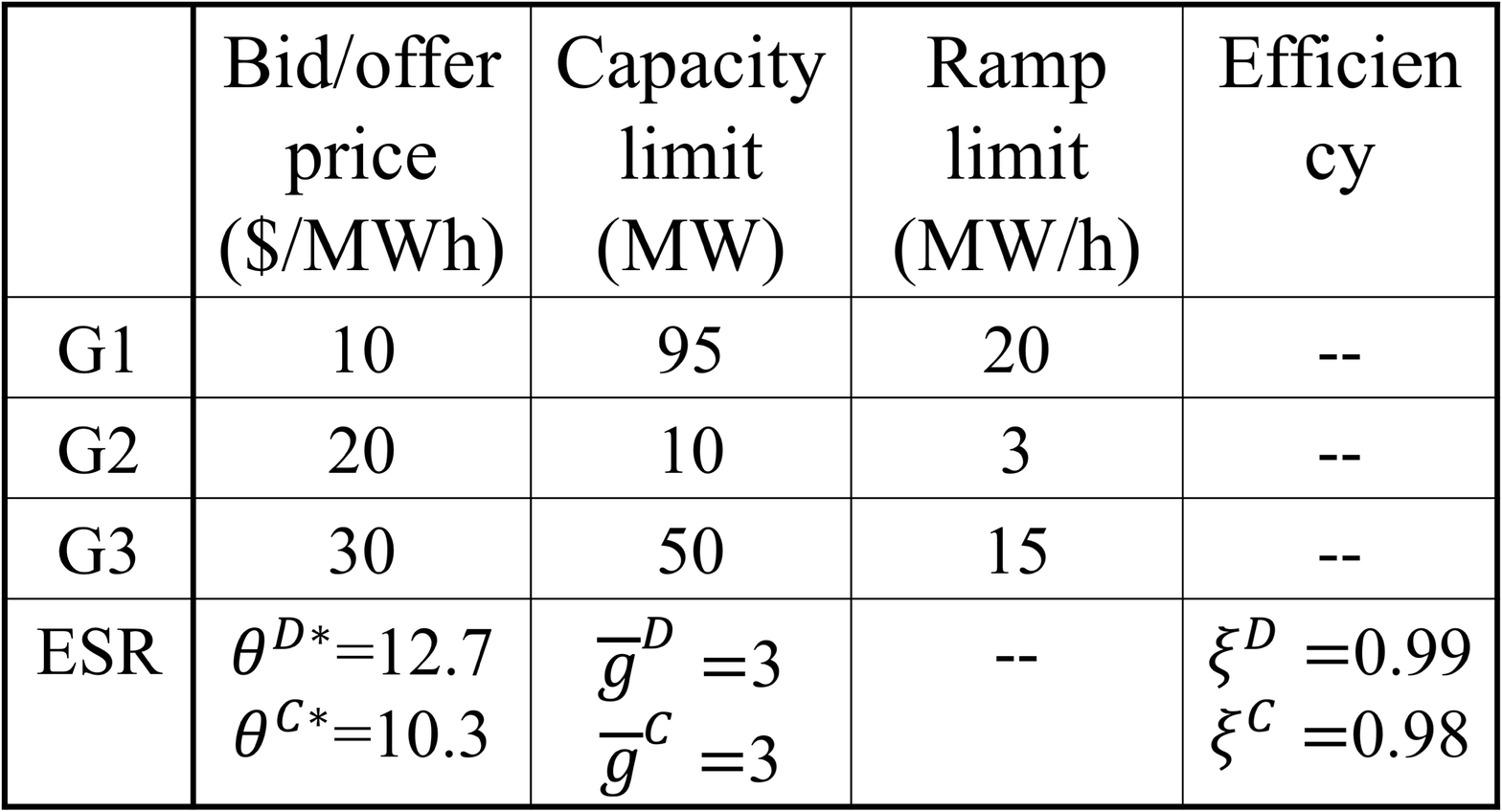}\scalefig{0.22}\epsfbox{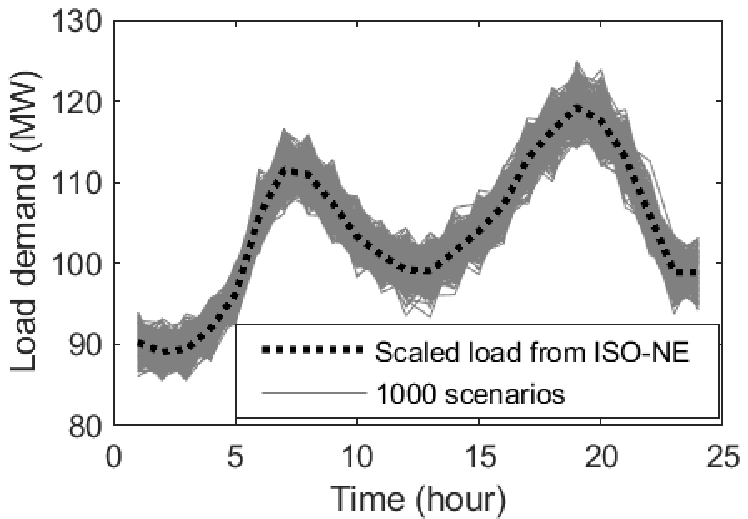}
\end{psfrags}
\vspace{-1em}\caption{\small Left: Parameter Settings. Right: demand traces.}
\label{fig:ParameterDemand}
\end{figure}

The left panel of Fig~\ref{fig:ParameterDemand}  shows the parameters of the generators and ESR in the simulations. Additionally, the initial SOC was at its lower limit $\underline{E}$ of 0.1 MWh. The upper limit $\bar{E}$ was at 25 MWh. Linear bid curves were adopted, and we evaluated the performance of TLMP and LMP, under varying load forecast errors. The right panel of Fig~\ref{fig:ParameterDemand} shows the 1000 realizations over 24 hour period generated from a scaled ISO New England (ISO-NE) load profile \cite{ISONE:21} with a standard deviation (STD) $5\%$ of the mean value.   

We used a standard forecasting error model\footnote{The forecast $\hat{d}_{(t+\tau)|t}$ at $t$  of demand $d_{t+\tau}$ is $\hat{d}_{(t+\tau)|t}=d_{t+\tau}+\sum_{i=1}^\tau \epsilon_\tau$ where $\epsilon_\tau$ is the realization of i.i.d. Gaussian with zero mean and variance $(\sigma d_{t+\tau})^2$.} inside each rolling-window, where the demand forecast $\hat{d}_{(t+\tau)|t}$ of $d_{t+\tau}$  at time $t$ had error variance $\tau(\sigma d_{t+\tau})^2$ with $\sigma=0.1\%,0.5\%,1\%,3\%$ representing varying load forecast errors. All simulations were conducted with rolling-window optimization over the 24-hour scheduling period, represented by 24 time intervals. In each rolling window optimization, the window size was 4 intervals, and $K=300$ load forecasting scenarios were considered  with $\epsilon_k=1/K, \forall k $ in the scenario-based stochastic optimization.

\subsection{Dispatch-following incentives}
Dispatch-following incentives were measured by the LOC in equation (\ref{eq:uplift}). Larger LOC payment represents higher incentive to deviate from the dispatch signal (in the absence of LOC). The top panels of Fig.~\ref{fig:metrics} shows LOC payment from the ISO to generators and ESR in different cases under different load forecast errors. As shown in Theorem~\ref{thm:StorageUniformPricingLOC}\& \ref{thm:R-TLMPLOC}, the LOC under TLMP was strictly zero. In contrast, the LOC under LMP was positive in all cases for all generators and ESR. Notice also that, under LMP, LOC for generators and ESR in stochastic rolling-window dispatch were lower than that in the deterministic dispatch. 

{\scriptsize
\begin{figure}[h]
\center
\begin{psfrags}
\scalefig{0.49}\epsfbox{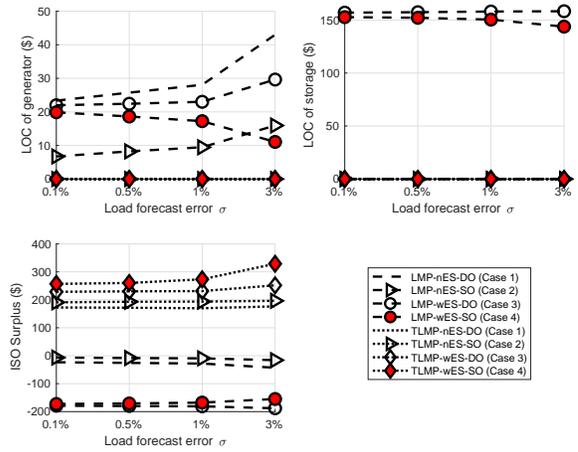}
\end{psfrags}
\vspace{-1em}
\caption{\scriptsize Top left: Generators' LOC vs. load forecast error. Top Right: ESR's LOC vs. load forecast error. Bottom left: ISO surplus vs. load forecast error.}
 \label{fig:metrics}
\end{figure}
}



\subsection{Revenue adequacy of ISO}
The bottom left panel of Fig.~\ref{fig:metrics} shows ISO's merchandising  surplus. Positive LOC payment to the market participants resulted in a negative merchandising surplus for ISO under LMP in all cases. ISO typically distributes deficits to demands as a financially neutral regulated utility, any deficit or surplus was redistributed to the consumers in a revenue reconciliation process \cite{Schweppe&Caramanis&Tabors&Bohn:88book}. For TLMP, the ramping price and SOC price led to a positive merchandising surplus in the average performance over 1000 scenarios. Coupled with the fact that TLMP always had zero LOC, TLMP resulted in a positive merchandising surplus for ISO.

With ESR participation, ISO has a higher positive surplus under TLMP than that without ESR. And by comparing deterministic and stochastic rolling-window dispatch, we observed that, under the stochastic rolling-window dispatch, ISO had larger surplus under TLMP, and less deficit under LMP.

{\scriptsize
\begin{figure}[h]
\center
\begin{psfrags}
\scalefig{0.35}\epsfbox{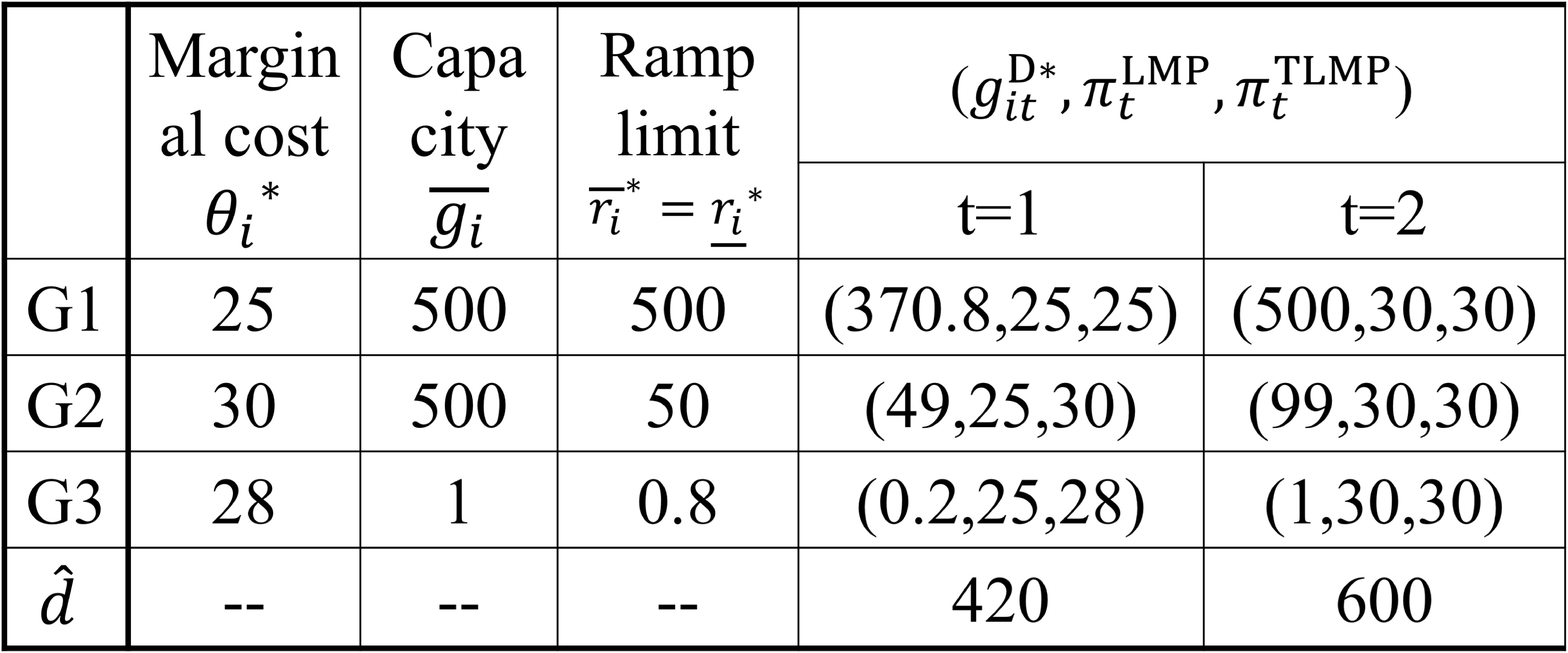}
\scalefig{0.24}\epsfbox{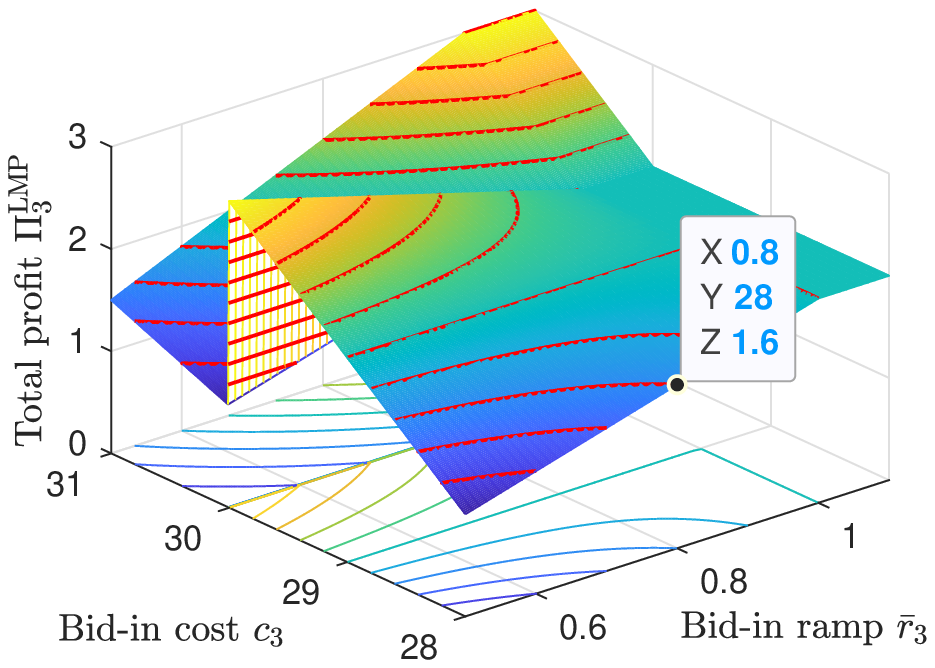}\scalefig{0.24}\epsfbox{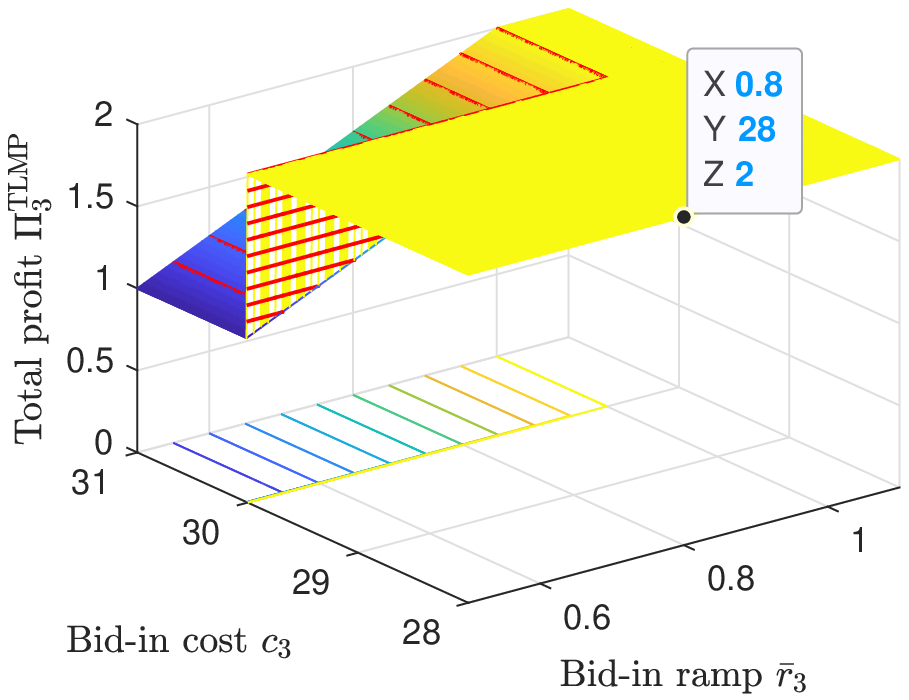}
\end{psfrags}
\vspace{-1em}
\caption{\scriptsize Top: Results at truthful bidding point. Bottom left: Expected total profit of G3 under R-LMP vs. $(c_3,\bar{r}_3)$. Bottom right: Expected total profit of G3 under R-TLMP vs. $(c_3,\bar{r}_3)$.}
 \label{fig:BidIncentive}
\end{figure}
}
\subsection{An example for untruthful-bidding incentive}\label{sec:UntruBidExample}
Here is an example that a generator with linear cost will have more profits by untruthfully revealing marginal cost $c$ and ramp limit $\bar{r}$ in rolling-window dispatch under R-LMP with LOC compensation. 

Following parameter settings from the top of Fig.~\ref{fig:BidIncentive}, which is similar to Table IV of \cite{Guo&Chen&Tong:21TPS}, a case of three generators and two rolling-windows with horizon $W=2$ was considered in this example. There was no load forecast error and the forecasted demands were $\hat{d}_{t=1}=(420, 600)~ {\rm MW}$ and $\hat{d}_{t=2}=(600, 600) ~{\rm MW}$.  The top of Fig.~\ref{fig:BidIncentive} \footnote{Unit for capacity is MW, for price is \$/MWh, for generation is MW, for ramping is MW/h}  lists rolling-window dispatch and pricing results at the truthful bidding point $\cbf^*=(25~30~28) ~{\rm \$/MWh},\bar{\rbf}^*=(500~50~0.8) ~{\rm MW/h}$. In this case, G3 would receive in-market profit \$1.4 and LOC \$0.2, thus the total profit is \$1.6.  

With a small generation capacity, G3 was to mimic a price taker who assumed unable to affect the market clearing price. Other generators were assumed to bid truthfully. The bottom of  Fig.~\ref{fig:BidIncentive} shows the {\em anticipated} total profit of G3 vs. $(c_3,\bar{r}_3)$ based on (\ref{eq:StoragePi_i}) and (\ref{eq:StoragePi_i_LMP}). It can be seen on the left panel of Fig.~\ref{fig:BidIncentive} that G3 anticipated receiving more profit if it were to withhold the bid-in ramp limit and increase bid-in cost under LMP. However, from the right panel of Fig.~\ref{fig:BidIncentive}, we observed that, locally, there was no profit gain by deviating from the truthful revelation point under TLMP. Notice the {\em anticipated} profit was calculated from the perspective of a price taker using $\pibf^{\mbox{\rm \tiny LMP}}=(25~30) $ \$/MWh and $\pibf^{\mbox{\rm \tiny TLMP}}_3=(28~30) $ \$/MWh.

\section{Conclusions}\label{sec:Conclusion}
Pricing stochastic operation in real-time is an open problem of increasing importance with greater stochasticity of real-time operation as the result of large-scale integration of renewable and broader participation of utility-scale storage and DERA from the distribution grid.  This work represents one of a few attempts in tackling the pricing problem for stochastic rolling-window dispatch.  

By establishing that uniform pricing cannot provide dispatch-following incentives, this work hopes to shift the attention to whether discrimination should be imposed outside the market clearing process via uplift payments or directly within the real-time market clearing process through nonuniform pricing such as TLMP. It is somewhat surprising that, as a direct generalization of LMP, TLMP removes the need for uplifts completely, independent of the quality of the demand forecasts.

A key point raised in this work is the truthful bidding aspects of competing pricing solutions. While one expects that price-taking generators would bid truthfully with marginal costs, the addition of LOC uplifts to generators’ profit calculation distorts the underlying argument for truthful bidding. With out-of-the-market payments, although price-taking generators do not assume their ability to affect the market-clearing price in constructing their bids and offers, they can manipulate biding parameters to influence their out-of-the-market payments. We demonstrate the possibility of such manipulations using a simple example.


\section{Appendix}\label{sec:Appendix}
\subsection{Proof of Lemma~\ref{lemma:SOC}}
With KKT conditions of the stochastic rolling-window dispatch in Fig.~\ref{fig:StochasticOptimization}, we have
\beq \label{eq:SOCprice}
\begin{array}{lrl}
-\underline{\delta}_{it}^*+\bar{\delta}_{it}^*+\phi_{it}^*-\sum_{k=1}^K\phi_{i(t+1)k}^*&=&0,\\
-\underline{\delta}_{i(t+1)k}^*+\bar{\delta}_{i(t+1)k}^*+\phi_{i(t+1)k}^*-\phi_{i(t+2)k}^*&=&0, \forall k,  \\
...&&\\
-\underline{\delta}_{i(t+W-1)k}^*+\bar{\delta}_{i(t+W-1)k}^*+\phi_{i(t+W-1)k}^*&=&0, \forall k, 
\end{array}
\eeq
which can be reached by taking derivative of the Lagrangian function over  SOC variables $E_{it},E_{i(t+1)k},...,E_{i(t+W-1)k}$. By adding all equations in (\ref{eq:SOCprice}), we have
\[\phi^*_{it}=\Delta \delta_{it}^*+\sum_{t'=t+1}^{t+W-1}\sum_{k=1}^K\Delta \delta_{it'k}^*,\]
where $\Delta \delta_{it}^*:=\underline{\delta}_{it}^*-\bar{\delta}_{it}^*, \Delta \delta_{it'k}^*:=\underline{\delta}_{it'k}^*-\bar{\delta}_{it'k}^*$.

\subsection{Proof of Corollary~\ref{corr:TLMP=LMP}}
From condition 1), dual variables for ramping constraints have $\Delta_{it}^{\mbox{\tiny C}*}=0, \Delta_{it}^{\mbox{\tiny D}*}=0$. 

From condition 2), dual variables for SoC limit constraints have $\bar{\delta}_{it}^*=0,\underline{\delta}_{it}^*=0$, $ \bar{\delta}_{it'k}^*=0, \underline{\delta}_{it'k}^*=0, \forall   t'\in \Hmsc_t \setminus  t,$ ~~~ $\forall k \in \{1,...,K\}$. 

By Lemma~\ref{lemma:SOC}, $\phi^*_{it}=\Delta \delta_{it}^*+\sum_{t'=t+1}^{t+W-1}\sum_{k=1}^K\Delta \delta_{it'k}^*=0$. 

So $\pi_{it}^{\mbox{\tiny TLMP}}=\pi^{\mbox{\rm\tiny LMP}}_{t} = \lambda^*_t$.

\subsection{Proof of Theorem~\ref{thm:StorageUniformPricingLOC}}
Suppose, under some realizations of stochastic demands and probabilistic load forecasts, the realization of the dispatch $(g_{it}^{\mbox{\tiny C}*},g_{it}^{\mbox{\tiny D}*})$ and  $(g_{jt}^{\mbox{\tiny C}*},g_{jt}^{\mbox{\tiny D}*})$ are optimal for (\ref{eq:Q}). Then, for all $t$, KKT conditions below should be satisfied: 

\beq \label{eq:StorageLOCKKT}
\begin{array}{l}
\frac{d}{d g} f^{\mbox{\tiny D}}_{kt}(g_{kt}^{\mbox{\tiny D}*})-\pi_t+1/\xi_k^{\mbox{\tiny D}}\psi^*_{kt}-\chi_{kt}^{{\mbox{\tiny D}}*} + \Delta\zeta^{{\mbox{\tiny D}}*}_{kt} =0,\forall k\in\{i,j\},\\
-\frac{d}{d g}f^{\mbox{\tiny C}}_{kt}(g_{kt}^{\mbox{\tiny C}*}) +\pi_t-\xi_k^{\mbox{\tiny C}}\psi^*_{kt}-\chi_{kt}^{{\mbox{\tiny C}}*}+ \Delta\zeta^{{\mbox{\tiny C}}*}_{kt} =0,\forall k\in\{i,j\},
\end{array}
\eeq
where $\chi_{kt}^{\mbox{\tiny D}*}:=\Delta \eta_{k(t+1)}^{{\mbox{\tiny D}}*}-\Delta \eta_{kt}^{{\mbox{\tiny D}}*}$, $\Delta \eta_{kt}^{\mbox{\tiny D}*}:=\bar{\eta}_{kt}^{\mbox{\tiny D}*}-\underline{\eta}_{kt}^{\mbox{\tiny D}*}$, $\Delta\zeta^{\mbox{\tiny D}*}_{kt}=\bar{\zeta}^{\mbox{\tiny D}*}_{kt} - \underline{\zeta}^{\mbox{\tiny D}*}_{kt}$. The same definition works for variables with superscript $C$.

It's known from the stochastic rolling window dispatch in Fig.~\ref{fig:StochasticOptimization} that $g^{\mbox{\tiny D}*}_{kt}g^{\mbox{\tiny C}*}_{kt}=0, \forall k, \forall t$. For ESR $i$ and $j$ fulfilling condition 2) in Theorem~\ref{thm:StorageUniformPricingLOC}, the nonzero charging/discharging power won't reach capacity and ramping limits at $t^*$. Meanwhile, SOC won't reach limits from $t^*$ to $T$. Here we show the contradiction when $g_{it^*}^{\mbox{\tiny D}*}  \in (0,\bar{g}^{\mbox{\tiny D}}_i)$ and $g_{jt^*}^{\mbox{\tiny D}*} \in (0,\bar{g}^{\mbox{\tiny D}}_j)$. Condition 2) gives $\Delta\zeta^{{\mbox{\tiny D}*}}_{kt^*}=\psi_{kt^*}^*=\chi_{it}^{{\mbox{\tiny D}}*} =0, \forall k\in\{i,j\}$. Under the uniform price $\pi_{t^*}$, (\ref{eq:StorageLOCKKT}) gives:
\[
 \pi_{t^*}= \frac{d}{d g} f^{\mbox{\tiny D}}_{it^*}(g_{it^*}^{\mbox{\tiny D}*}) = \frac{d}{d g} f^{\mbox{\tiny D}}_{jt^*}(g_{jt^*}^{\mbox{\tiny D}*}). 
  \]
This contradicts condition 1). And we can reach similar contradiction when $g_{it^*}^{\mbox{\tiny D}*}  \in (0,\bar{g}^{\mbox{\tiny D}}_i)$ and $g_{jt^*}^{\mbox{\tiny C}*} \in (0,\bar{g}^{\mbox{\tiny C}}_j)$ , $g_{it^*}^{\mbox{\tiny C}*}  \in (0,\bar{g}^{\mbox{\tiny C}}_i)$ and $g_{jt^*}^{\mbox{\tiny D}*} \in (0,\bar{g}^{\mbox{\tiny D}}_j)$, and $g_{it^*}^{\mbox{\tiny C}*}  \in (0,\bar{g}^{\mbox{\tiny C}}_i)$ and $g_{jt^*}^{\mbox{\tiny C}*} \in (0,\bar{g}^{\mbox{\tiny C}}_j)$. So there does not exist a uniform pricing scheme under which both ESR $i$ and $j$ have optimal self-scheduling plans at the rolling-window dispatch signals. And nonzero LOC is needed to compensate ESR when conditions in Theorem~\ref{thm:StorageUniformPricingLOC} are fulfilled. 

Because the stochastic inelastic demand process and the stochastic demand forecast process have continuous distributions, all conditions 1) and 2) still hold when the realizations slightly deviate. So under the given conditions, at least one of ESR $i$ and $j$ has nonzero LOC with positive probability, and uniform pricing cannot provide dispatch-following incentive support in this situation. 

\subsection{Proof of Theorem~\ref{thm:R-TLMPLOC}}
Solve stochastic rolling-window dispatch in Fig.~\ref{fig:StochasticOptimization} under any realizations of the stochastic inelastic demand process and the stochastic demand forecast process, we have rolling-window dispatch signals $(\gbf^{\mbox{\tiny D}*},\gbf^{\mbox{\tiny C}*})$, which comes from the binding interval of each rolling-window optimization. Therefore, the KKT condition of rolling-window dispatch signals gives 
\beq \label{eq:DispatchKKTSO}
\begin{array}{l}
\frac{d}{d g}f^{\mbox{\tiny D}}_{it}(g_{it}^{\mbox{\tiny D}*})- \lambda_t +1/\xi_i^{\mbox{\tiny D}} \phi_{it}^* -\Delta_{it}^{\mbox{\tiny D}*}+ \Delta\rho^{{\mbox{\tiny D}}*}_{it}= 0,\forall i,t,\\
-\frac{d}{d g} f^{\mbox{\tiny C}}_{it}(g_{it}^{\mbox{\tiny  C}*})+ \lambda_t  -\xi_i^{\mbox{\tiny C}} \phi_{it}^*-\Delta_{it}^{\mbox{\tiny C}*}+ \Delta\rho^{{\mbox{\tiny C}}*}_{it}= 0,\forall i,t,
\end{array}
\eeq
where $\Delta_{it}^{\mbox{\tiny D}*}:=\sum_{ k=1}^ { K}\Delta \mu_{i(t+1)k}^{{\mbox{\tiny D}}*}-\Delta \mu_{it}^{{\mbox{\tiny D}}*}$,$\Delta \mu_{i(t+1)k}^{\mbox{\tiny D}*}:=\bar{\mu}_{i(t+1)k}^{\mbox{\tiny D}*}-\underline{\mu}_{i(t+1)k}^{\mbox{\tiny D}*}$, $\Delta \mu_{it}^{\mbox{\tiny D}*}:=\bar{\mu}_{it}^{\mbox{\tiny D}*}-\underline{\mu}_{it}^{\mbox{\tiny D}*}$, $\Delta\rho^{\mbox{\tiny D}*}_{it}:=\bar{\rho}^{\mbox{\tiny D}*}_{it} - \underline{\rho}^{\mbox{\tiny D}*}_{it}$. The same definition works for variables with superscript $C$. 
Notice (\ref{eq:DispatchKKTSO}) comes from the binding interval of each rolling-window optimization rather than a single multi-interval dispatch problem.

For the self-optimal dispatch, it must satisfy (\ref{eq:StorageLOCKKTindivi}), KKT conditions of (\ref{eq:Q}):
\beq \label{eq:StorageLOCKKTindivi}
\begin{array}{l}
\frac{d }{d g}f_{it}^{\mbox{\tiny D}}(p^{{\mbox{\tiny D}}*}_{it})-\pi_t+1/\xi_i^{\mbox{\tiny D}}\psi^*_{it}-\chi_{it}^{{\mbox{\tiny D}}*} + \Delta\zeta^{{\mbox{\tiny D}}*}_{it} =0,\forall i,t,\\
-\frac{d }{d g} f_{it}^{\mbox{\tiny C}}(p^{{\mbox{\tiny C}}*}_{it})+\pi_t-\xi_i^{\mbox{\tiny C}}\psi^*_{it}-\chi_{it}^{{\mbox{\tiny C}}*}+ \Delta\zeta^{{\mbox{\tiny C}}*}_{it} =0,\forall i,t,
\end{array}
\eeq
where $\chi_{it}^{\mbox{\tiny D}*}:=\Delta \eta_{i(t+1)}^{{\mbox{\tiny D}}*}-\Delta \eta_{it}^{{\mbox{\tiny D}}*}$, $\Delta \eta_{it}^{\mbox{\tiny D}*}:=\bar{\eta}_{it}^{\mbox{\tiny D}*}-\underline{\eta}_{it}^{\mbox{\tiny D}*}$, $\Delta\zeta^{\mbox{\tiny D}*}_{it}=\bar{\zeta}^{\mbox{\tiny D}*}_{it} - \underline{\zeta}^{\mbox{\tiny D}*}_{it}$. The same definition works for variables with superscript $C$.

The KKT conditions of the individual optimization shown in (\ref{eq:StorageLOCKKTindivi}) can be satisfied like equation (\ref{eq:DispatchKKTSO}) by setting $g^{\mbox{\tiny D}*}_{it}= p^{{\mbox{\tiny D}}*}_{it},g^{\mbox{\tiny C}*}_{it}= p^{{\mbox{\tiny C}}*}_{it},\Delta\rho^{\mbox{\tiny D}*}_{it}=\Delta\zeta^{\mbox{\tiny D}*}_{it},\Delta\rho^{\mbox{\tiny C}*}_{it}=\Delta\zeta^{\mbox{\tiny C}*}_{it},\bar{\eta}_{it}^{\mbox{\tiny D}*}=\underline{\eta}^{\mbox{\tiny D}*}_{it}=\bar{\eta}_{it}^{\mbox{\tiny C}*}=\underline{\eta}^{\mbox{\tiny C}*}_{it}=\psi^*_{it}=0,\forall i,t$. That way, we have $\mbox{\rm LOC}(\pibf^{\mbox{\rm\tiny TLMP}},\gbf_i^{\mbox{\tiny D}*},\gbf_i^{\mbox{\tiny C}*})=0$ from equation (\ref{eq:uplift}).

For every realization of stochastic demands and probabilistic load forecasts, the above proof can show that LOC equal to 0, i.e. LOC equals to 0 almost surely. So TLMP can support dispatch-following incentives of ESRs in stochastic storage operation with zero LOC.

\subsection{Proof of Theorem~\ref{thm:TLMPBidLOCESR}}
Under any realizations of the stochastic inelastic demand process and the stochastic demand forecast process, it's known from Theorem \ref{thm:R-TLMPLOC} that TLMP can support dispatch-following incentives of ESRs with zero LOC. So, from (\ref{eq:StoragePi_i}), the profit maximization problem for the optimal bidding strategy under TLMP is \footnote{The ESR index $i$ is dropped here for brevity} 
\beq \label{eq:pi}
\begin{array}{lcl}
&\underset{\{\thetabf\}}{\rm maximize}  & (\pibf^{\mbox{\tiny TLMP-D}})^{\T}\gbf^{\mbox{\tiny D}*}(\thetabf)-\sum_{t=1}^T q^{\mbox{\tiny D}}_{t}(g^{\mbox{\tiny D}*}_{t}(\thetabf))\\
&&+\sum_{t=1}^T q^{\mbox{\tiny C}}_{t}(g^{\mbox{\tiny C}*}_{t}(\thetabf))-(\pibf^{\mbox{\tiny TLMP-C}})^{\T}\gbf^{\mbox{\tiny C}*}(\thetabf)  \\[0.5em]
& {\rm subject~to}&\\
& & -\underline{\rbf}_i^{\mbox{\tiny D}*}\le \Abf \gbf^{\mbox{\tiny D}*}(\thetabf) \le \bar{\rbf}_{i}^{\mbox{\tiny D}*}, \\
&& -\underline{\rbf}_i^{\mbox{\tiny C}*}\le \Abf \gbf^{\mbox{\tiny C}*}(\thetabf) \le \bar{\rbf}_{i}^{\mbox{\tiny C}*}, \\
& & \underline{\Ebf}^*\le \Ebf^{*}(\thetabf) \le \bar{\Ebf}^*, \\
 &&{\bf 0}\leq \gbf^{\mbox{\tiny D}*}(\thetabf) \leq\bar{\gbf}^{\mbox{\tiny D}*},\\
 &&{\bf 0}\leq \gbf^{\mbox{\tiny C}*}(\thetabf) \leq \bar{\gbf}^{\mbox{\tiny C}*},
\end{array} \hfill
\eeq
where the bidding parameter $\thetabf$ includes operational parameters such as bid-in cost, ramping limits, SOC limits and generation capacity limits, i.e. $\thetabf:=(c_i^{\mbox{\tiny D}},c_i^{\mbox{\tiny C}},\bar{r}_i^{\mbox{\tiny D}},\bar{r}_i^{\mbox{\tiny C}},\underline{r}_i^{\mbox{\tiny D}},\underline{r}_i^{\mbox{\tiny C}},\bar{E}_i,\underline{E}_i,\bar{g}_i^{\mbox{\tiny D}},\bar{g}_i^{\mbox{\tiny C}},\underline{g}_i^{\mbox{\tiny D}},\underline{g}_i^{\mbox{\tiny C}})$. Correspondingly, the truthful bidding parameter is $\thetabf^*=(c_i^{\mbox{\tiny D}*},c_i^{\mbox{\tiny C}*},\bar{r}_i^{\mbox{\tiny D}*},\bar{r}_i^{\mbox{\tiny C}*},\underline{r}_i^{\mbox{\tiny D}*},\underline{r}_i^{\mbox{\tiny C}*},\bar{E}_i^*,\underline{E}_i^*,\bar{g}_i^{\mbox{\tiny D}*},\bar{g}_i^{\mbox{\tiny C}*},\underline{g}_i^{\mbox{\tiny D}*},\underline{g}_i^{\mbox{\tiny C}*})$ with truthful components. 

Constraints in eq.(\ref{eq:pi}) descirbe the truthful feasible dispatch region, as an ESR expects the rolling-window dispatch signal to fall into its truthful dispatch region; or else, ESR won't be able to follow the dispatch signal. 

With the price-taker assumption, an ESR takes the given TLMP $\pibf^{\mbox{\rm\tiny TLMP}}$ as fixed parameters. As $q_{t}^{\mbox{\tiny D}}(\cdot)=f_{t}^{\mbox{\tiny D}}(\cdot|\thetabf^*)$ by definition, we know that the optimal solution of (\ref{eq:pi}) equals to $Q(\cdot|\thetabf^*)$ in (\ref{eq:Q}). It's known from the dispatch following incentive in Theorem \ref{thm:R-TLMPLOC} that  $(\gbf^{\mbox{\tiny D}*}(\thetabf^*),\gbf^{\mbox{\tiny C}*}(\thetabf^*))$ gives optimal value $Q(\cdot|\thetabf^*)$ in (\ref{eq:Q}), so
\beq \label{eq:StoragePi_TLMPALL}
\begin{array}{l}
\Pi^{\mbox{\tiny TLMP}}(\thetabf^*) = Q(\cdot|\thetabf^*) = (\pibf^{\mbox{\tiny TLMP-D}})^{\T}\gbf^{\mbox{\tiny D}*}(\thetabf^*)-\sum_{t=1}^T q^{\mbox{\tiny D}}_{t}(g^{\mbox{\tiny D}*}_{t}(\thetabf^*))\\
~~~~+\sum_{t=1}^T q^{\mbox{\tiny C}}_{t}(g^{\mbox{\tiny C}*}_{t}(\thetabf^*))-(\pibf^{\mbox{\tiny TLMP-C}})^{\T}\gbf^{\mbox{\tiny C}*}(\thetabf^*) \ge (\pibf^{\mbox{\tiny TLMP-D}})^{\T}\gbf^{\mbox{\tiny D}}\\
~~~~-\sum_{t=1}^T q^{\mbox{\tiny D}}_{t}(g^{\mbox{\tiny D}}_{t})+\sum_{t=1}^T q^{\mbox{\tiny C}}_{t}(g^{\mbox{\tiny C}}_{t})-(\pibf^{\mbox{\tiny TLMP-C}})^{\T}\gbf^{\mbox{\tiny C}},
\end{array}
\eeq
for every $\gbf^{\mbox{\tiny D}},\gbf^{\mbox{\tiny C}}$ in the truthful dispatch region.  And it's obvious that a price-taker's bid under TLMP can only influence dispatch $\gbf^{\mbox{\tiny D}*}(\thetabf)$ and $\gbf^{\mbox{\tiny C}*}(\thetabf)$ in (\ref{eq:pi}).  So we have $\Pi(\thetabf^*) \ge  \Pi(\thetabf), \forall \thetabf$.  So, a price-taking ESR has no incentive to deviate from the truthful bidding point $\thetabf^*$.

For every realization of the stochastic inelastic demand process and the stochastic demand forecast process, the above proof works. So, it is optimal for a price-taking ESR to bid truthfully with its truthful parameter $\thetabf^*$ under all realizations of stochastic demands and probabilistic load forecasts. \hfil

{
\bibliographystyle{IEEEtran}
\bibliography{BIB}
}

\end{document}